\documentclass[10pt,conference]{IEEEtran}
\usepackage{setspace}
\usepackage{xcolor,soul,framed} 
\usepackage[noadjust]{cite}

\colorlet{shadecolor}{yellow}

\usepackage[pdftex]{graphicx}
\graphicspath{{../pdf/}{../jpeg/}}
\DeclareGraphicsExtensions{.pdf,.jpeg,.png}
\usepackage[caption=false, font=footnotesize]{subfig}
\usepackage{amssymb,mathtools}
\usepackage{array}
\usepackage{mdwmath}
\usepackage{mdwtab}
\usepackage{eqparbox}
\usepackage{url}
\usepackage[ruled,vlined,linesnumbered]{algorithm2e}
\usepackage{enumitem}
\usepackage[mathcal]{euscript}
\usepackage{tabularx} 
\usepackage{multirow} 
\usepackage{booktabs} 
\usepackage{makecell} 
\usepackage{aligned-overset}
\usepackage{url}
\usepackage{multicol} 
\usepackage{balance}
\usepackage{bbm}
\usepackage{amsmath}
\usepackage{amsthm}
\usepackage{stfloats}
\usepackage{pifont}

\graphicspath{{./Figures/}}

\renewcommand{\vec}[1]{\boldsymbol{\mathrm{#1}}}

\newcommand{\TRIS}{\mathtt{TRIS}}
\newcommand{\ARIS}{\mathtt{ARIS}}
\newcommand{\RIS}{\mathtt{R}}

\newcommand{\user}{\mathtt{UE}}

\newcommand{\BS}{\mathtt{BS}}

\newcommand{\pl}{\mathrm{PL}}

\usepackage[hidelinks]{hyperref}
\usepackage{xcolor}
 \usepackage[left=.65in,                               right=.65in,                     top=1in,  bottom=1.1in]{geometry}

\begin{document}
\bstctlcite{IEEEexample:BSTcontrol}

\title{Hierarchical RIS Association for FR3 Space-Air-Ground Integrated Networks}

\author{
  \IEEEauthorblockN{Muddasir Rahim~(\textit{Member,~IEEE}), Irfan Azam~(\textit{Member,~IEEE}), and Soumaya Cherkaoui~(\textit{Senior~Member,~IEEE})}
  \IEEEauthorblockA{Department of Computer Engineering and Software Engineering, Polytechnique Montreal, Canada} \IEEEauthorblockA{Emails: muddasir.rahim@polymtl.ca, irfan.azam@polymtl.ca, soumaya.cherkaoui@polymtl.ca}
}
\maketitle
\begin{abstract}
The growing demand for diverse sixth-generation (6G) applications requires efficient connectivity, seamless service continuity, and extended network coverage. Although terrestrial infrastructures can effectively serve most wireless user equipment (UE), communication reliability may be significantly degraded in remote or blocked environments. To address this challenge, this paper proposes a reconfigurable intelligent surface (RIS)-assisted space-air-ground integrated network (SAGIN) operating in the upper mid-band (FR3) spectrum. The considered network consists of a terrestrial base station (BS), a low Earth orbit (LEO) satellite, multiple terrestrial RISs (TRISs), and unmanned aerial vehicle-mounted RISs (ARISs), where the BS serves as the primary access node, while the LEO satellite provides complementary connectivity when terrestrial links are unavailable or cannot satisfy the required quality-of-service (QoS) requirements. Within this framework, we formulate a hierarchical RIS association problem to maximize the achievable network sum rate under RIS capacity constraints. The resulting optimization problem is a mixed-integer nonconvex problem due to the binary association variables and the coupling among UEs competing for limited RIS resources. To efficiently solve the problem, we develop a hierarchical many-to-one matching-based RIS association framework. Specifically, UEs with satisfactory terrestrial connectivity are first associated with TRISs, whereas UEs that cannot be reliably served by the terrestrial network are subsequently associated with ARISs through the non-terrestrial network. Simulation results demonstrate that the proposed hierarchical association framework significantly outperforms greedy and random association schemes in terms of the achievable network sum rate.
\end{abstract}
\begin{IEEEkeywords}
RI), FR3, SAGIN
\end{IEEEkeywords}
\section{INTRODUCTION}\label{introduction}
\IEEEPARstart{T}{he} evolution toward sixth-generation (6G) wireless networks is driven by the need to support a wide range of emerging applications with higher data rates, ubiquitous connectivity, extended coverage, and reliable communication services. These requirements are particularly challenging in environments with limited terrestrial infrastructure, such as remote areas, rural regions, disaster recovery scenarios, and highly dynamic communication environments~\cite{6GIoT}. To address these challenges, the upper mid-band (UMB), also known as frequency range 3 (FR3), has recently emerged as a promising spectrum candidate for future 6G networks~\cite{rahim2026dual}. In particular, the 7--15 GHz band provides an attractive tradeoff between the limited bandwidth of sub-6 GHz systems and the severe propagation losses encountered in millimeter-wave and terahertz bands  \cite{chaves2024coverage, bjornson2025enabling}. Consequently, FR3 is well suited for supporting next-generation wireless services that require both enhanced capacity and wide-area coverage.

 Despite these advantages, ensuring reliable communication in FR3 networks remains challenging due to increased path loss, penetration loss, and sensitivity to line-of-sight (LoS) blockage compared with lower-frequency cellular bands, particularly in dense urban, industrial, and remote environments \cite{cui20236g,khan2022urllc, rahim2026ris}. Reconfigurable intelligent surfaces (RISs) have emerged as a promising solution to overcome these propagation limitations by intelligently reconfiguring the wireless environment through passive beamforming, thereby enhancing coverage without requiring active relaying~\cite{huang2019reconfigurable}. Recent studies have investigated RIS-assisted beamforming, user association, and resource management for different wireless systems \cite{rahim2024multi, kara2024reconfigurable,rahim2026reliable}. However, most existing works focus on terrestrial networks (TNs), while RIS-assisted space-air-ground integrated networks (SAGINs) operating in the FR3 spectrum remain largely unexplored. In particular, efficient hierarchical user equipment (UE)-RIS association across terrestrial and aerial RISs is still an open research problem.

Although terrestrial RISs (TRISs) can significantly improve local coverage, they may not fully resolve connectivity issues when both the direct BS-UE link and the TRIS-assisted path are severely degraded or blocked. Such situations frequently arise in dense urban, industrial, and remote environments, where obstacles block LoS propagation. To enhance service continuity, SAGINs have emerged as a promising architecture that combines terrestrial and non-terrestrial communication segments to provide ubiquitous connectivity \cite{azari2022evolution}. In such networks, terrestrial BSs offer low-cost and low-latency access whenever reliable terrestrial links are available, whereas low Earth orbit (LEO) satellites provide complementary coverage when TN cannot satisfy the required quality-of-service (QoS) requirements. Furthermore, integrating RISs into SAGINs further improves communication reliability by reconfiguring the wireless propagation environment through passive beamforming. In particular, TRISs enhance local terrestrial coverage, while aerial RISs (ARISs) provide flexible assistance for non-terrestrial network (NTN) \cite{you2022enabling, mohsan2023irs}.

Although several recent studies have investigated hierarchical RIS-assisted TN and NTN \cite{tian2023hybrid, rahim2025hierarchical,RIS_STIN}, most existing works focus on terahertz or very-high-frequency systems, where direct terrestrial links are often assumed to be unavailable. This assumption does not accurately reflect FR3 networks, in which direct BS-UE links, TRIS-assisted links, and LEO-assisted ARIS links may all provide viable communication opportunities depending on the channel conditions. Motivated by this observation, this paper proposes a hierarchical RIS association framework for FR3 SAGINs. In the proposed system, the base station (BS) serves as the primary access node, while UEs with satisfactory terrestrial connectivity are first associated with TRISs. When the terrestrial link is unavailable or fails to satisfy the required QoS, the remaining UEs are associated with the ARIS through the LEO-assisted communication mode. Accordingly, a hierarchical deferred acceptance-based many-to-one matching framework is developed to efficiently determine the UE associations while maximizing the network sum rate.

\subsection{Contributions}\label{contributions}
Existing RIS association schemes primarily focus on TNs and do not account for the hierarchical architecture and heterogeneous communication links of SAGINs. In such networks, user association becomes significantly more challenging due to the coexistence of terrestrial and non-terrestrial access, multi-tier RIS deployments, and limited RIS capacities. Motivated by these challenges, this paper develops a hierarchical RIS association framework for RIS-assisted SAGINs operating in the FR3 spectrum. The main contributions of this paper are summarized as follows:
\begin{itemize}
  \item We propose a novel RIS-assisted SAGIN architecture for FR3 networks comprising a terrestrial BS, a LEO satellite, multiple TRISs, and ARISs. In the proposed architecture, the BS serves as the primary access node, while the LEO satellite provides complementary connectivity when terrestrial links are unavailable or unable to satisfy the required QoS requirements.
  \item We formulate a hierarchical RIS association problem to maximize the achievable network sum rate under user association and RIS capacity constraints. To efficiently solve the formulated problem, we propose a hierarchical many-to-one matching-based RIS association framework. Specifically, users with satisfactory terrestrial connectivity are first associated with TRISs, whereas the remaining users are subsequently associated with ARISs through the NTN.
  \item Simulation results demonstrate that the proposed hierarchical RIS association framework significantly outperforming greedy search (GS) and random search (RS) schemes in terms of the achievable network sum rate.
\end{itemize}
\section{System Model and Problem Formulation} \label{model}
\begin{figure}[!t]
     \centering
\includegraphics[width=0.8\linewidth]{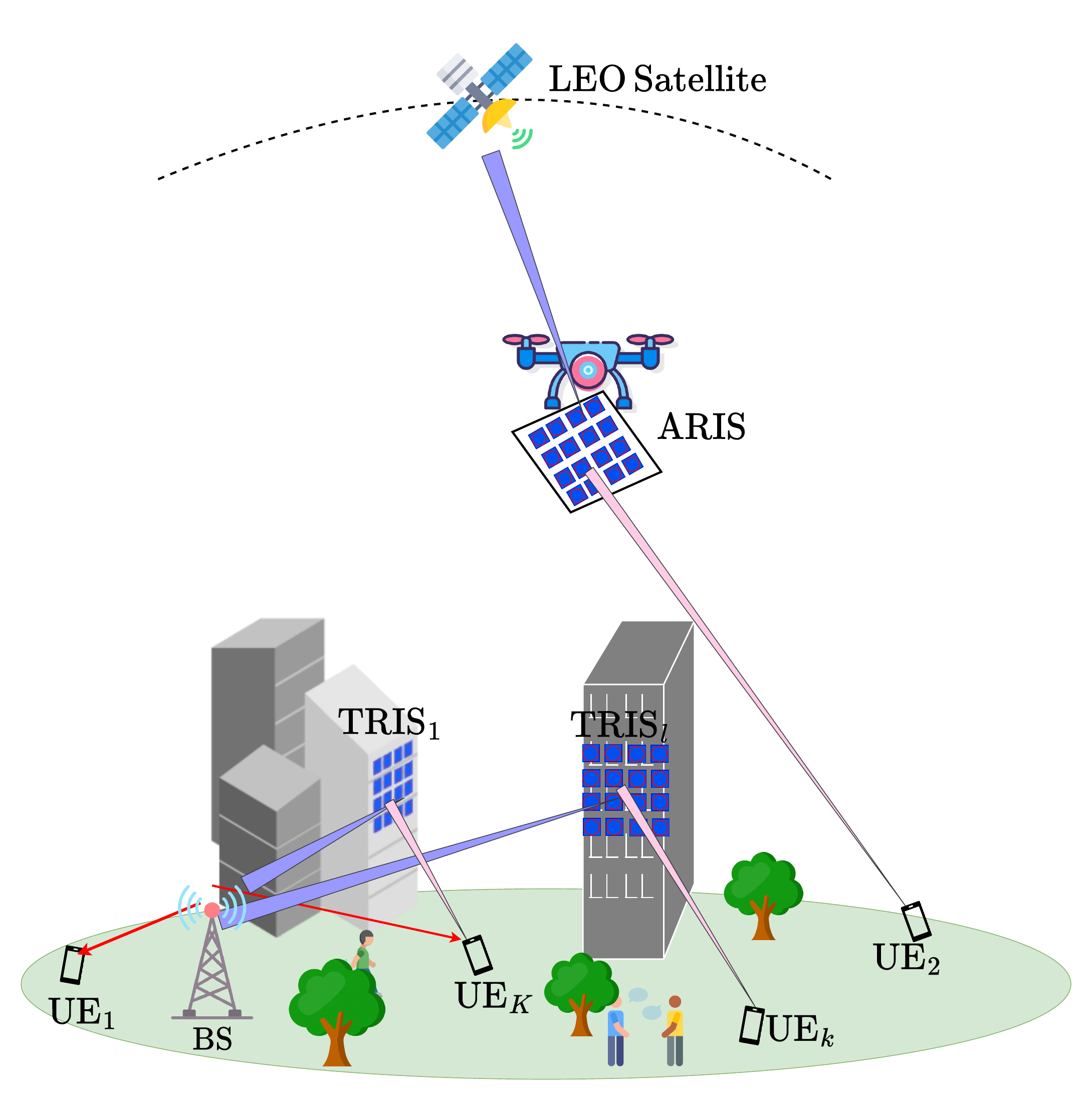}
    \caption{System model of the proposed RIS-assisted SAGIN operating in the FR3 spectrum.
     }
     \label{m1}   
\end{figure}

We consider SAGIN comprising a terrestrial BS, a LEO satellite, ARIS, multiple TRISs, and $K$ set of user equipment (UEs), as illustrated in Fig.~\ref{m1}. The BS, equipped with $N$ number of antennas, serves as the primary access node due to its low deployment and operational cost. Specifically, UEs are preferentially served through the terrestrial network whenever a direct BS-to-UE link or TRIS-assisted link is available. However, when the BS link is blocked or unable to satisfy the required QoS constraints, communication is maintained through the LEO satellite via the ARIS layer. In this manner, the proposed architecture combines the cost-effectiveness of TNs with the ubiquitous coverage and resilience of NTNs. To enhance coverage and mitigate blockage effects, the network is supported by a set of $L$ TRISs and a single ARIS. The TRIS set is denoted by $\mathcal{L}=\TRIS_1,\TRIS_2,\ldots,\TRIS_l,\ldots,\TRIS{L}$. Each RIS consists of $M=M_x \times M_y$ passive reflecting elements arranged in a uniform planar array, where $M_x$ and $M_y$ denote the numbers of reflecting elements along the horizontal and vertical directions, respectively. In the Cartesian coordinate system, the center locations of the BS, $k^{th}$ UE ($\user_{k}$), $l^{th}$ TRIS ($\RIS_{l}$), ARIS, and LEO satellite are represented as ${\mathbf p_{b} = (x_{b}, y_b, z_{b})}$, ${\mathbf p_{k} = (x_{k}, y_{k}, z_{k})}$, ${\mathbf p_{l} = (x_{l}, y_{l}, z_{l})}$, ${\mathbf p_{a} = (x_{a}, y_{a}, z_{a})}$, and ${\mathbf p_{s} = (x_{s}, y_{s}, z_{s})}$ respectively.
\subsection{Channel Modeling of the Proposed SAGIN}
We consider an RIS-assisted SAGIN, where each UE is served either through the terrestrial BS or the LEO satellite based on the adopted hierarchical association strategy. Specifically, the BS acts as the primary access node due to its low latency and cost-effective deployment. When the terrestrial link is unavailable or cannot satisfy the required QoS requirements, the corresponding UE is served through the LEO-assisted communication link. The TN is assisted by multiple TRISs, whereas the NTN employs an ARIS to enhance the communication quality. Let $\mathbf{h}_{b,k}\in\mathbb{C}^{N}$ denote the direct channel vector between the BS and $\user_k$, which is expressed as
 \begin{align}\label{eq1}
  \textstyle  [h_{b,k}]_n=h_{(b,n),k} = \sqrt{\pl_{(b,n),k}} \, e^{-j \omega r_{a,l}},
 \end{align}
where $\pl_{(b,n),k}$ is the pathloss from $\BS_n$ to $\user_{k}$, $r_{b,k}$ is the distance between the BS and $\user_{k}$, and $\omega = \frac{2\pi f}{c}$ is the wave number at frequency $f$. The channel matrix between the BS and $\TRIS_l$ is denoted by $\vec{H}_{b,l}\in\mathbb{C}^{M\times N}$ and can be expressed as
\begin{align}\label{channel_ap_tirs}
\textstyle [\vec{H}_{b,l}]_{n,m}
=
h_{(b,n),(l,m)}
=
\sqrt{\pl_{(b,n),(l,m)}}
\,e^{-j\omega r_{b,l}},
\end{align}
where $\pl_{(b,n),(l,m)}$ represents the path loss between the $n^{th}$ BS antenna and the $m^{th}$ reflecting element of $\TRIS_l$ and $r_{b,l}$ is the distance between the BS and $\TRIS_{l}$. Furthermore, the channel vector from $\TRIS_l$ to $\user_k$ is denoted by $\vec{h}_{l,k}\in\mathbb{C}^{M}$ and is given by
\begin{align}\label{channel_tirs_user}
\textstyle [\vec{h}_{l,k}]_m
=
h_{(l,m),k}
=
\sqrt{\pl_{(l,m),k}}
\,e^{-j\omega r_{l,k}},
\end{align}
where $\pl_{(l,m),k}$ denotes the path loss from the $m^{th}$ element of $\TRIS_t$ to $\user_k$ and $r_{l,k}$ is the distance between $\TRIS_{l}$ and $\user_k$. Let $\vec{G_{s,a}}\in\mathbb{C}^{M\times N}$ denotes the channel matrix from the LEO satellite to $\ARIS$, which can be expressed as
\begin{align}\label{channel-L-U}
\textstyle[\vec{G}_{s,a}]_{n,m}
=
g_{(s,n),(a,m)}
=
\sqrt{\pl_{(s,n),(a,m)}}
\,e^{-j\omega r_{s,a}},
\end{align}
where $\pl_{(s,n),(a,m)}$ represents the path loss between the $n^{th}$ satellite antenna and the $m^{th}$ reflecting element of $\ARIS$ and $r_{s,a}$ is the distance between the LEO satellite and $\ARIS$.
Furthermore, the channel vector from $\ARIS$ to $\user_k$ is denoted by $\vec{g}_{a,k}\in\mathbb{C}^{M}$ and is given by
\begin{align}\label{channel_uirs_user}
\textstyle[\vec{g}_{a,k}]_m
=
g_{(a,m),k}
=
\sqrt{\pl_{(a,m),k}}
\,e^{-j\omega r_{a,k}},
\end{align}
where $\pl_{(a,m),k}$ denotes the path loss from the $m^{th}$ element of $\ARIS$ to $\user_k$ and $r_{a,k}$ is the distance between $\ARIS$ and $\user_k$. Let $\gamma_{k,l}\in\{0,1\}$ denote the UE-TRIS pairing variable, where $\gamma_{k,l}=1$ if $\user_k$ is paired with $\TRIS_l$, and $0$ otherwise. Similarly, let $\psi_{k}\in\{0,1\}$ denote the UE-ARIS pairing variable, where $\psi_{k}=1$ if $\user_k$ is associated with the $\ARIS$, and $0$ otherwise. Moreover, $\alpha_k\in\{0,1\}$ denote the transmission mode selection variable. To prioritize terrestrial communication while using the LEO-assisted link as a fallback, $\alpha_k$ is determined based on the QoS of the terrestrial link. Specifically, let $\widehat{R}_k$ denote the achievable data rate of $\user_k$ over the BS-assisted link, and let $R_{\min}$ denote the minimum required data rate. Then, $\alpha_k=1$ if $\widehat{R}_k\geq R_{\min}$, and $\alpha_k=0$ otherwise, where $\alpha_k=1$ indicates that $\user_k$ is served by the BS, whereas $\alpha_k=0$ indicates that the $\user_k$ is served through the LEO-assisted link. Finally, the effective downlink channel of $\user_k$ can be expressed as
\begin{align}\label{eq:effective_channel}
\vec{h}_k\!=\!\alpha_k(\vec{h}_{b,k}\!\!+\!\!\sum_{l=1}^{L}\!\!
\gamma_{k,l} \vec{H}_{b,l}^{\sf H}\vec{\Theta}\vec{h}_{l,k})
\!+\!(1\!-\!\alpha_k) ( \psi_{k} \vec{G}_{s,a}^{\sf H}\vec{\Theta} \vec{h}_{a,k}),
\end{align}
where $\Theta$ denotes the RIS reflection matrix, which is assumed to be preconfigured with ideal phase shifts to maximize the reflected signal power, and the reflection amplitude of each RIS element is assumed to be unity.
Based on the effective channel in~\eqref{eq:effective_channel}, the signal-to-interference-and-noise ratio (SINR) for $\user_k$ can be written as
\begin{equation}\label{sinr}
\text{SINR}_{k} = \frac{p_{k}|(\mathbf{h}_{k})^H {\vec{w}}_k|^2}{\sum_{i \neq k}^{K}p_i|(\mathbf{h}_{k})^H {\vec{w}}_i|^2 + \sigma^2},
\end{equation}
where $p_k$ and ${\vec{w}}_k$ are the transmit power scaling factor and the beamforming vector for $\user_k$, respectively. Moreover, $n_k$ denotes the additive white Gaussian noise (AWGN) with zero mean and variance $\sigma^2$. Finally, the achievable data rate for $\user_k$ can be written as
\begin{align}\label{rate}
   R_{k} =\log_2(1 + \frac{p_{k}|(\mathbf{h}_{k})^H {\vec{w}}_k|^2}{\sum_{i \neq k}^{K}p_i|(\mathbf{h}_{k})^H {\vec{w}}_i|^2 + \sigma^2}).
\end{align}
\subsection{Optimization Problem Formulation}\label{OP}
In the considered SAGIN, the achievable data rate of each UE depends on the selected communication mode and the associated RIS. Specifically, UEs with satisfactory terrestrial connectivity are served by the terrestrial BS through TRISs, whereas the remaining UEs are served by the LEO satellite through ARISs. Consequently, the objective is to determine the hierarchical UE-RIS association, including the UE-TRIS and UE-ARIS associations, to maximize the achievable network sum rate 
\begin{figure*}[!tb]
     \centering
\includegraphics[width=0.7\linewidth]{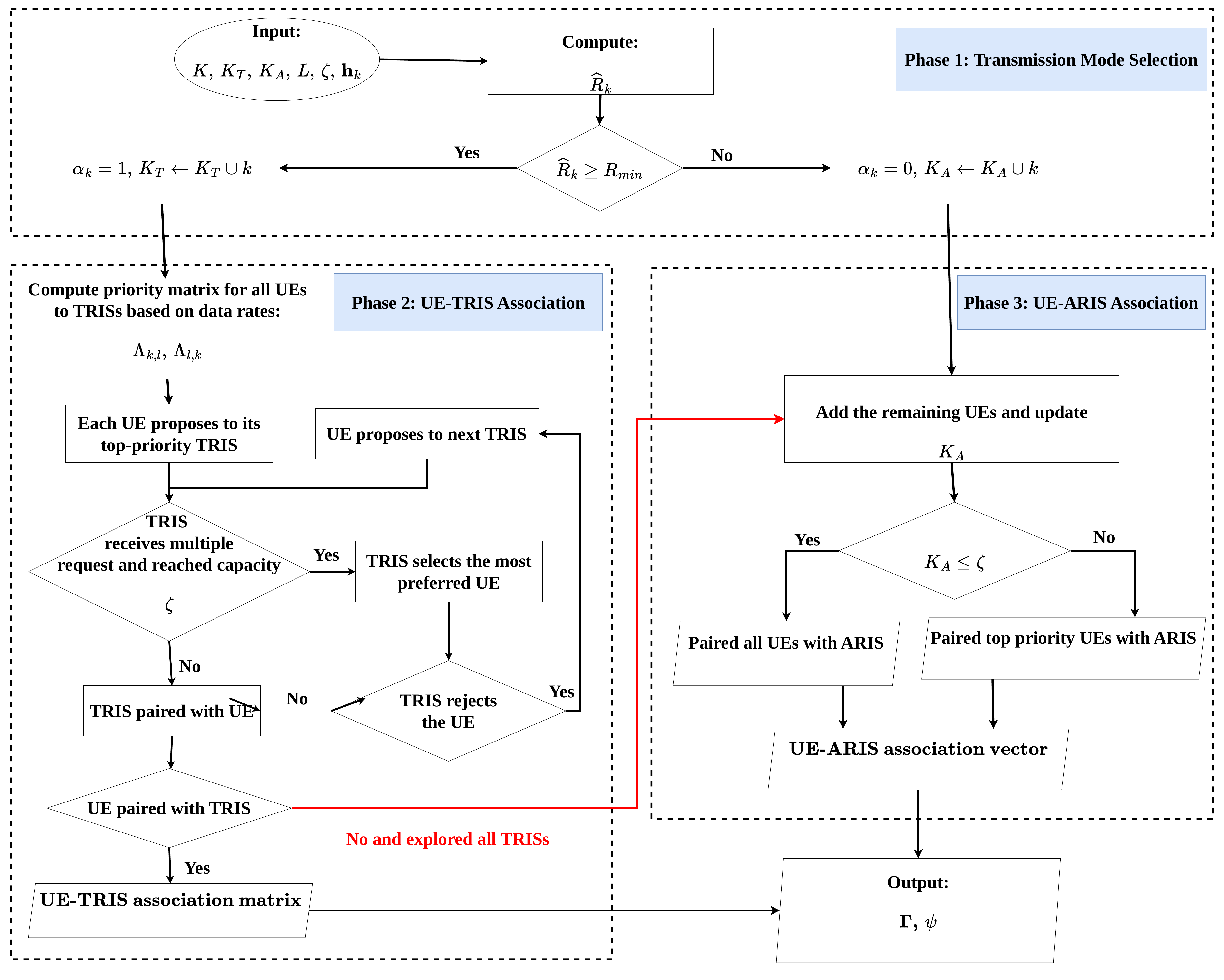}
     \caption{Flowchart of the proposed hierarchical deferred acceptance-based many-to-one matching framework for UE association in the considered FR3 SAGIN.}
     \label{flow}   
\end{figure*}
\begin{subequations}\label{eq_opt_prob}
\begin{alignat}{2}
&\mathcal{P}: \underset{\vec{\Gamma, \, \psi}}{\text{ maximize}}
&
&\quad\sum_{k=1}^K R_k (\vec{ \Gamma, \psi}), \label{eq_optProb}\\ 
&\text{subject to} 
 &&\mathcal{C}1:\quad  \gamma_{k,l}\in\{0,1\}, \ \forall k\in K,\, l\in L, \label{eq_constraint3}\\
 &&&\mathcal{C}2:\quad  \psi_{k}\in\{0,1\}, \ \forall k\in K, \label{eq_constraint4}\\
 &&&\mathcal{C}3:\quad  \textstyle\sum_{l=1}^{L} \gamma_{k,l} \le \alpha_k,\ \forall k\in K, \label{eq_constraint5}\\ 
 &&&\mathcal{C}4:\quad \textstyle\sum_{k=1}^{K} \gamma_{k,t} \le \zeta,\ \forall l\in L, \label{eq_constraint6}\\
 &&&\mathcal{C}5:\quad  \textstyle \psi_{k} \le (1-\alpha_k),\ \forall k\in K, \label{eq_constraint7}\\ 
 &&&\mathcal{C}6:\quad \textstyle\sum_{k=1}^{K} \psi_{k} \le \zeta,\, \label{eq_constraint8}
\end{alignat}
\end{subequations}
where $\zeta$ denotes the maximum numbers of UEs that can be supported by each RIS. The constraints can be interpreted as
follows:
\begin{itemize}
    \item $\mathcal{C}1$ and $\mathcal{C}2$: Impose binary assignment decisions.
    \item $\mathcal{C}3$: Ensures that only UEs assigned to the BS-aided mode can be associated with a TRIS
    \item $\mathcal{C}4$: Enforces the maximum capacity of each TRIS.
    \item $\mathcal{C}5$:  Ensures that only UEs assigned to the LEO-aided mode can be associated with an ARIS.
    \item $\mathcal{C}6$: Enforces the maximum capacity of each ARIS.
\end{itemize}
\section{Proposed Deferred Acceptance Framework}\label{proposed}
Problem~\eqref{eq_opt_prob} is a mixed-integer nonconvex optimization problem due to the binary UE pairing variables. In addition, the achievable data rate in~\eqref{rate} depends on the selected communication mode and the associated RIS, resulting in coupled association decisions among UEs competing for limited RIS resources. Consequently, the pairing variables are highly interdependent, since assigning a UE to a particular RIS directly affects the association opportunities and achievable data rates of the remaining UEs. As a result, the formulated problem reduces to a combinatorial optimization problem that is closely related to the class of NP-hard user association and link selection problems. Therefore, we propose a hierarchical deferred acceptance many-to-one matching framework to efficiently solve the formulated UE association problem. Fig.~\ref{flow} illustrates the proposed hierarchical many-to-one matching framework based on the deferred acceptance algorithm for UE-RIS association. The framework consists of three sequential phases, namely transmission mode selection, UE-TRIS association, and UE-ARIS association. In the first phase, the transmission mode is determined for each UE according to the achievable data rate of the terrestrial BS-assisted link. Specifically, the achievable data rate $\widehat{R}_k$ is computed and compared with the minimum data rate $R_{\min}$. If $\widehat{R}_k \geq R_{\min}$, the UE is assigned to the terrestrial communication mode and added to the terrestrial user set $K_T$. Otherwise, the UE is assigned to the LEO-assisted communication mode and added to the aerial user set $K_A$.

In the second phase, the UEs in $K_T$ are associated with the available TRISs through a hierarchical deferred acceptance many-to-one matching process. First, each UE computes its preference list by ranking the TRISs according to the achievable data rate. Subsequently, each UE proposes to its most preferred TRIS. If the number of received requests does not exceed the TRIS capacity, the requesting UEs are accepted. Otherwise, the TRIS retains the most preferred UEs according to its preference list and rejects the remaining requests. The rejected UEs subsequently propose to their next preferred TRIS, and the process continues until every UE is associated or all TRISs have been explored.

Finally, in the third phase, all UEs that were initially assigned to the non-terrestrial mode together with the UEs rejected by all TRISs are added to the aerial user set. If the number of remaining UEs does not exceed the ARIS capacity, all UEs are associated with the ARIS. Otherwise, the ARIS admits the highest priority UEs according to the priority list while the remaining requests are rejected. The algorithm then outputs the final UE-TRIS and UE-ARIS association matrices, denoted by $\mathbf{\Gamma}$ and $\mathbf{\psi}$, respectively.
\section{Simulation Results and Discussion}\label{sim}
In this section, the performance of the proposed hierarchical RIS association framework is evaluated through Monte Carlo simulations. The considered FR3 SAGIN consists of a single terrestrial BS equipped with $N=64$ antennas and $K=50$. Each RIS comprises $100\times100$ passive reflecting elements arranged in a uniform planar array. The side length of reflecting element is $\lambda/2 $, where $\lambda$ denotes the carrier wavelength. The system operates in the FR3 spectrum at a carrier frequency of $15$ GHz with a total bandwidth of $200$ MHz. Unless otherwise specified, the BS transmit power is fixed at $30$ dBm and is equally distributed among all served UEs. The thermal noise power spectral density is set to -$174$ dBm/Hz, and the receiver noise figure is assumed to be $10$ dB. The BS-UE and LEO-assisted channels follow a Rician fading model with a Rician factor of 10. The considered deployment area is a square region of size $500\times500 m^2$. The BS is located at the center of the coverage area with a height of $20 m$, while the LEO satellite is positioned at an altitude of $400 km$. Furthermore, the TRISs are deployed at heights between $20 m$ and $30 m$, whereas the ARIS operates at an altitude between $200 m$ and $400 m$. All simulations are implemented in MATLAB, and the simulation results are averaged over $10^6$ independent network realizations with randomly generated UE locations and channel conditions to ensure statistical reliability.  To evaluate the effectiveness of the proposed hierarchical RIS association framework, the performance is compared with the following benchmark schemes. \textbf{Greedy Search (GS):} In this scheme, each UE is sequentially associated with the RIS that provides the highest achievable data rate. If the preferred TRIS has reached its maximum serving capacity, the UE is assigned to the next preferred TRIS. When no TRIS can accommodate the UE or the terrestrial link cannot satisfy the required QoS requirements, the UE is associated with the ARIS through the LEO link, subject to its capacity constraint. \textbf{Random Search (RS):} In this benchmark, each UE is randomly associated with an available TRIS or the ARIS without considering channel conditions or achievable data rates.

\begin{figure}[!htp]
     \centering
\includegraphics[width=\linewidth]{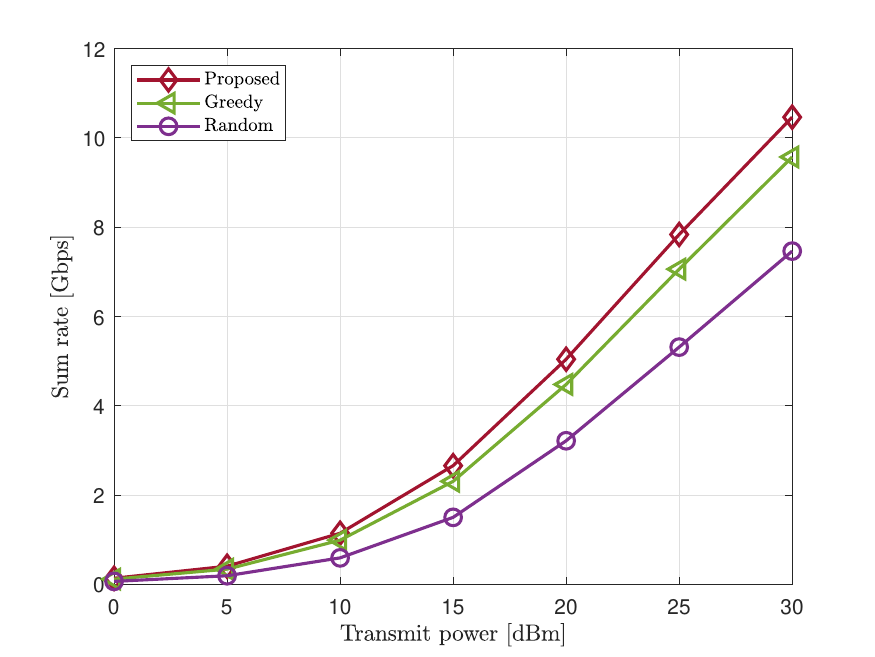}
     \caption{Network sum rate versus the transmit power budget.}
     \label{rate_power}   
\end{figure}

\begin{figure}[!htp]
     \centering
\includegraphics[width=\linewidth]{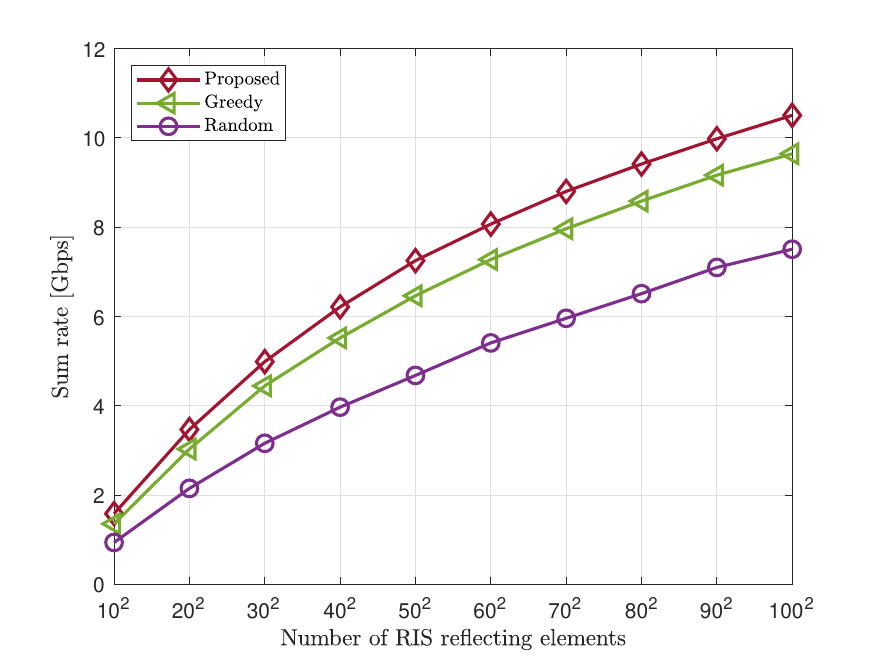}
     \caption{Network sum rate versus number of RIS reflecting elements.}
     \label{rate_reflectors}   
\end{figure}

Fig.~\ref{rate_power} compares the network sum rate of the proposed scheme with the GS and RS schemes as the BS transmit power increases. The network sum rate of all schemes increases with the transmit power due to the strong received SINR. Furthermore, the proposed scheme consistently achieves the highest network sum rate across the entire transmit power range. This performance gain is attributed to the proposed hierarchical many-to-one matching strategy, which efficiently associates UEs with the most suitable TRISs or the ARIS while satisfying the RIS capacity constraints. In contrast, the GS scheme makes locally optimal association decisions that may lead to suboptimal resource utilization, whereas the RS scheme ignores channel conditions entirely, resulting in the lowest achievable network sum rate.
Fig.~\ref{rate_reflectors} illustrates the network sum rate versus the number of RIS reflecting elements. As the number of reflecting elements increases, the network sum rate improves for all schemes due to the enhanced passive beamforming gain. Moreover, the proposed hierarchical RIS association framework consistently outperforms the GS and RS schemes by efficiently associating UEs with the most suitable RISs.

\begin{figure}[!htp]
     \centering
\includegraphics[width=\linewidth]{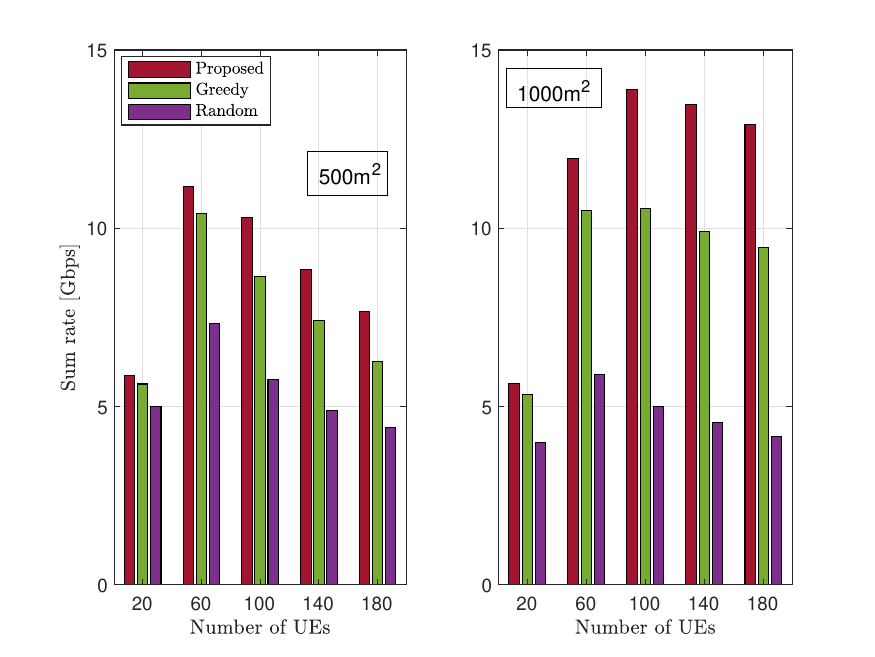}
     \caption{Network sum rate versus number of UEs.}
     \label{rate_user}   
\end{figure}

\begin{figure}[!htp]
     \centering
\includegraphics[width=0.8\linewidth]{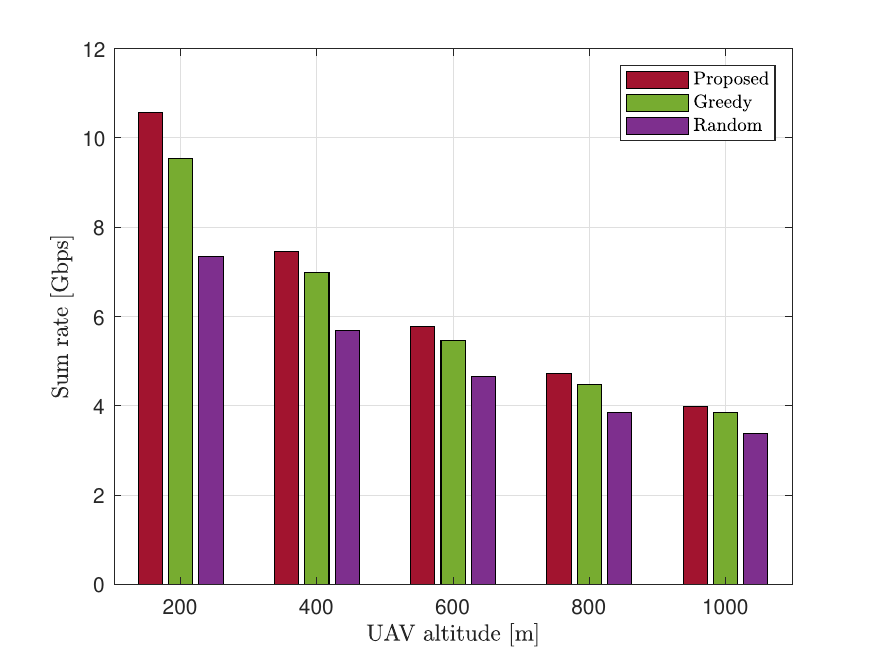}
     \caption{Network sum rate versus UAV altitude.}
     \label{rate_height}   
\end{figure}

Fig.~\ref{rate_user} compares the network sum rate versus the number of UEs for network areas of $500 \times 500 m^2$ and $1000 \times 1000 m^2$. For both deployment scenarios, the achievable network sum rate initially increases as more UEs provide greater multiuser diversity. However, beyond a certain number of UEs, the network sum rate gradually decreases due to increased competition for the limited RIS resources and stronger interference. Moreover, the larger network area achieves a higher network sum rate because it accommodates more UEs while benefiting from the proposed hierarchical RIS association framework. In both scenarios, the proposed scheme consistently outperforms the GS and RS schemes.
Fig.~\ref{rate_height} shows the network sum rate versus the ARIS altitude. As the ARIS altitude increases, the network sum rate gradually decreases for all schemes due to the increased propagation distance and the resulting path loss. Moreover, the proposed method consistently achieves the highest network sum rate by efficiently associating UEs with the available RIS resources, outperforming the GS and RS schemes across all considered altitudes.
\section{Conclusions}\label{conc}
This paper investigated the hierarchical RIS association problem in RIS-assisted SAGINs operating in the FR3 spectrum. A hierarchical many-to-one matching-based association framework is proposed to maximize the achievable network sum rate while satisfying RIS capacity constraints. Specifically, UEs with satisfactory terrestrial connectivity are associated with TRISs, whereas the remaining UEs are served through ARISs via the NTN, thereby ensuring seamless connectivity and efficient utilization of network resources. Simulation results demonstrated that the proposed hierarchical association framework consistently outperformed greedy and random association schemes in terms of the achievable network sum rate. Future work will extend the proposed framework by jointly optimizing RIS association with resource allocation and investigating robust designs under imperfect channel state information and dynamic network conditions.
\bibliographystyle{IEEEtran}
\balance
\bibliography{References}

@article{6GIoT,
  author={Nguyen, Dinh C. and Ding, Ming and Pathirana, Pubudu N. and Seneviratne, Aruna and Li, Jun and Niyato, Dusit and Dobre, Octavia and Poor, H. Vincent},
  journal={IEEE Internet of Things Journal}, 
  title={{6G Internet of Things: A Comprehensive Survey}}, 
  year={2022},
  volume={9},
  number={1},
  pages={359-383},
  doi={10.1109/JIOT.2021.3103320}}

@article{huang2019reconfigurable,
  title={{Reconfigurable intelligent surfaces for energy efficiency in wireless communication}},
  author={Huang, Chongwen and Zappone, Alessio and Alexandropoulos, George C and Debbah, M{\'e}rouane and Yuen, Chau},
  journal={IEEE Trans. Wireless Commun.},
  volume={18},
  number={8},
  pages={4157--4170},
  year={2019},
  publisher={IEEE}
}

@article{rahim2024multi,
  title={{Multi-IRS-aided terahertz networks: Channel modeling and user association with imperfect CSI}},
  author={Rahim, Muddasir and Nguyen, Thanh Luan and Kaddoum, Georges and Do, Tri Nhu},
  journal={IEEE Open J. Commun. Soc.	},
   volume={5},
  pages={836--855},
  year={2024},
  publisher={IEEE}
}

@article{tian2023hybrid,
  title={{Hybrid deployment of multi-hierarchical RISs for probabilistic LoS communication in dense urban}},
  author={Tian, Zhong and Wang, Jing and Chen, Zhengchuan and Wang, Min and Jia, Yunjian and Wu, Dapeng O},
  journal={IEEE Commun. Lett.	},
  volume={27},
  number={8},
  pages={2207--2211},
  year={2023},
  publisher={IEEE}
}

@article{azari2022evolution,
  title={{Evolution of non-terrestrial networks from 5G to 6G: A survey}},
  author={Azari, M Mahdi and Solanki, Sourabh and Chatzinotas, Symeon and Kodheli, Oltjon and Sallouha, Hazem and Colpaert, Achiel and Montoya, Jesus Fabian Mendoza and Pollin, Sofie and Haqiqatnejad, Alireza and Mostaani, Arsham and others},
  journal={IEEE Commun. Surv. Tutorials	},
  volume={24},
  number={4},
  pages={2633--2672},
  year={2022},
  publisher={IEEE}
}

@article{chaves2024coverage,
  title={{Coverage evaluation of 7--15 GHz bands from existing sites}},
  author={Chaves, F and Chizhik, D and Du, J and Ghosh, A and Love, B and Visotsky, E},
  journal={Nokia White Paper},
  year={2024}
}

@article{khan2022urllc,
  title={{URLLC and eMBB in 5G industrial IoT: A survey}},
  author={Khan, Benish Sharfeen and Jangsher, Sobia and Ahmed, Ashfaq and Al-Dweik, Arafat},
  journal={IEEE Open Journal of the Communications Society},
  volume={3},
  pages={1134--1163},
  year={2022},
  publisher={IEEE}
}

@article{bjornson2025enabling,
  title={{Enabling 6G performance in the upper mid-band by transitioning from massive to gigantic MIMO}},
  author={Bj{\"o}rnson, Emil and Kara, Ferdi and Kolomvakis, Nikolaos and Kosasih, Alva and Ramezani, Parisa and Salman, Murat Babek},
  journal={IEEE Open J. Commun. Soc.},
  year={2025},
  publisher={IEEE}
}

@article{rahim2026dual,
  title={{Dual-Tier IRS-Assisted Mid-Band 6G Mobile Networks: Robust Beamforming and User Association}},
  author={Rahim, Muddasir and Cherkaoui, Soumaya},
  journal={arXiv preprint arXiv:2602.00431},
  year={2026}
}

@article{rahim2026reliable,
  title={{Reliable IoT Communications in 6G Non-Terrestrial Networks with Dual RIS}},
  author={Rahim, Muddasir and Cherkaoui, Soumaya},
  journal={arXiv preprint arXiv:2602.00438},
  year={2026}
}

@inproceedings{cui20236g,
  title={6{G} Wireless Communications in 7–24 {GH}z Band: Opportunities, Techniques, and Challenges}, 
  author={Cui, Zhuangzhuang and Zhang, Peize and Pollin, Sofie},
  booktitle={2025 IEEE International Symposium on Dynamic Spectrum Access Networks (DySPAN)},
  pages={1--8},
  year={2025},
  organization={IEEE}
}

@article{you2022enabling,
  title={{Enabling smart reflection in integrated air-ground wireless network: IRS meets UAV}},
  author={You, Changsheng and Kang, Zhenyu and Zeng, Yong and Zhang, Rui},
  journal={IEEE Wireless Commun.	},
  volume={28},
  number={6},
  pages={138--144},
  year={2022},
  publisher={IEEE}
}

@article{kara2024reconfigurable,
  title={{Reconfigurable Intelligent Surfaces in Upper Mid-Band 6G Networks: Gain or Pain?}},
  author={Kara, Ferdi and Demir, {\"O}zlem Tu{\u{g}}fe and Bj{\"o}rnson, Emil},
  journal={arXiv preprint arXiv:2407.05754},
  year={2024}
}

@article{mohsan2023irs,
  title={{IRS-assisted UAV communications: A comprehensive review}},
  author={Mohsan, Syed Agha Hassnain and Li, Yanlong},
  journal={arXiv preprint arXiv:2306.15838},
  year={2023}
}

@article{rahim2025hierarchical,
  title={{Hierarchical IRS-assisted user association for sum-rate maximization in hybrid THz LEO-HAPS-terrestrial networks}},
  author={Rahim, Muddasir and Basit, Abdul and Selim, Bassant and Kaddoum, Georges},
  journal={IEEE Commun. Lett.},
  year={2025},
  publisher={IEEE}
}

@article{RIS_STIN,
  author={Li, Xin and Li, Yongjun and Song, Xinkang and Shao, Long and Li, Hai},
  journal={IEEE Photonics Journal}, 
  title={RIS Assisted UAV for Weather-Dependent Satellite Terrestrial Integrated Network With Hybrid FSO/RF Systems}, 
  year={2023},
  volume={15},
  number={5},
  pages={1-17}}

@article{rahim2026ris,
  title={{RIS-Assisted Joint Resource Allocation for 6G FR3 IoT Networks}},
  author={Rahim, Muddasir and Azam, Irfan and Cherkaoui, Soumaya},
  journal={arXiv preprint arXiv:2604.02487},
  year={2026}
}

\end{document}